\documentclass[final,5p,times,twocolumn]{elsarticle}
\usepackage{lineno,hyperref}
\usepackage{amssymb}
\usepackage{xcolor}
\usepackage{capt-of}
\usepackage{placeins}
\usepackage{stfloats}
\modulolinenumbers[5]
\journal{Nuclear Instruments and Methods in Physics Research A}

\begin{document}
\frenchspacing
\begin{frontmatter}
\title{Characterization of immersed SiPM arrays in liquid scintillator between room temperature and \texorpdfstring{$-30\,^{\circ}\mathrm{C}$}{-30 degrees C}}

\author[cau]{M.~S.~Kwak}

\author[cau]{J.~S.~Chung}

\author[cau]{C.~Ha\corref{cor}}
\ead{chha@cau.ac.kr}

\author[cau]{T.~Z.~Huang}

\author[cau]{J.~Kim}

\author[cau]{S.~A.~Kim}

\author[cau]{H.~Kimku\corref{cor}}
\ead{kimkuhani9@gmail.com}

\author[cau]{B.~C.~Koh}

\author[cup,ust]{H.~S.~Lee}

\author[cau]{S.~Lee}

\author[cau]{Y.~J.~Lee}

\author[cau]{J.~Seo}

\cortext[cor]{Corresponding author}
\address[cau]{Department of Physics, Chung-Ang University, Seoul 06974, Republic of Korea}
\address[cup]{Center for Underground Physics, Institute for Basic Science (IBS), Daejeon 34126, Republic of Korea}
\address[ust]{IBS School, University of Science and Technology (UST), Daejeon 34113, Republic of Korea}

\begin{abstract}
  Liquid scintillator detectors instrumented with distributed silicon photomultiplier (SiPM) arrays can be used in compact, topology-sensitive, and low-background experiments, but the temperature dependence of SiPMs immersed directly in the scintillation medium has not been widely characterized.
  We report the operation of a 125-liter linear-alkylbenzene-based liquid scintillator detector read out by 125 SiPM channels immersed in the active volume, over the range from room temperature to $-30\,^{\circ}\mathrm{C}$.
  The detector response was measured with cosmic-ray muons, including stopping muons followed by their Michel-electron decay. Cooling from $+15\,^{\circ}\mathrm{C}$ to $-30\,^{\circ}\mathrm{C}$ reduced the SiPM dark-count rate by a factor of 13.5, increased the single-photoelectron response by 49.1\%, and increased the cosmic-ray muon light yield by 16.7\%.
  The improved photoelectron separation and baseline stability at low temperature enabled a selection of stopping-muon events, from which the effective muon lifetime was measured to be $1959\pm132\,\mathrm{ns}$, consistent with the value expected for a hydrocarbon scintillator once $\mu^{-}$ capture on carbon is taken into account.
\end{abstract}


\end{frontmatter}

\section{Introduction}

Silicon photomultipliers (SiPMs) are increasingly used in liquid scintillator detectors because of their compact size, low operating voltage, high photon-detection efficiency, and insensitivity to magnetic fields~\cite{Piemonte2019}. Unlike conventional photomultiplier tubes, SiPMs can be assembled into compact tiles or distributed readout modules, as in the TAO and DUNE near-detector designs~\cite{TAO-CDR,DUNE:2021tad}.

Distributing SiPMs directly within the active scintillator volume can shorten the optical path between photon production and detection. The performance of immersed SiPMs depends strongly on temperature. Lowering the temperature suppresses thermally generated dark counts, which is beneficial for low-light scintillation measurements. It also changes the breakdown voltage, gain, and single-photoelectron (SPE) response. The scintillation light yield and detector trigger behavior can change as well~\cite{TAO-CDR,Acerbi2019,TAO-LS}. These effects are coupled, so the net response of a full detector cannot be inferred from individual-sensor data alone and must be measured \emph{in situ}.

At the single-sensor level the temperature dependence of SiPMs has been studied extensively, from cryogenic characterization for noble-liquid detectors~\cite{Baudis2018,Wang2021VUV4,Borden2024} to bench measurements spanning intermediate and room temperatures~\cite{Anfimov2021}. SiPM-based scintillator antineutrino detectors have also been built, but typically with the sensors coupled to the ends or faces of optically segmented scintillator elements rather than immersed in a monolithic volume~\cite{SANDD}. In contrast, the temperature-dependent response of a large array of SiPMs immersed directly throughout an organic liquid-scintillator volume has not been characterized. This response reflects the combined changes in sensor gain, dark-count rate, scintillation light yield, optical transport, and trigger behavior.

The Chung-Ang University Neutrino Detection Yolk (CANDY) program is developing a liquid-scintillator detector for short-baseline reactor antineutrino oscillation measurements. A distributed array of SiPMs immersed throughout the active volume reconstructs the event topology and position~\cite{Kimku2026Reconstruction}. Accurate event localization both suppresses backgrounds and, by fixing the antineutrino interaction point, determines the baseline $L$ that an oscillation analysis requires.

\begin{figure*}[!t]
  \begin{center}
      \includegraphics[width=0.85\textwidth]{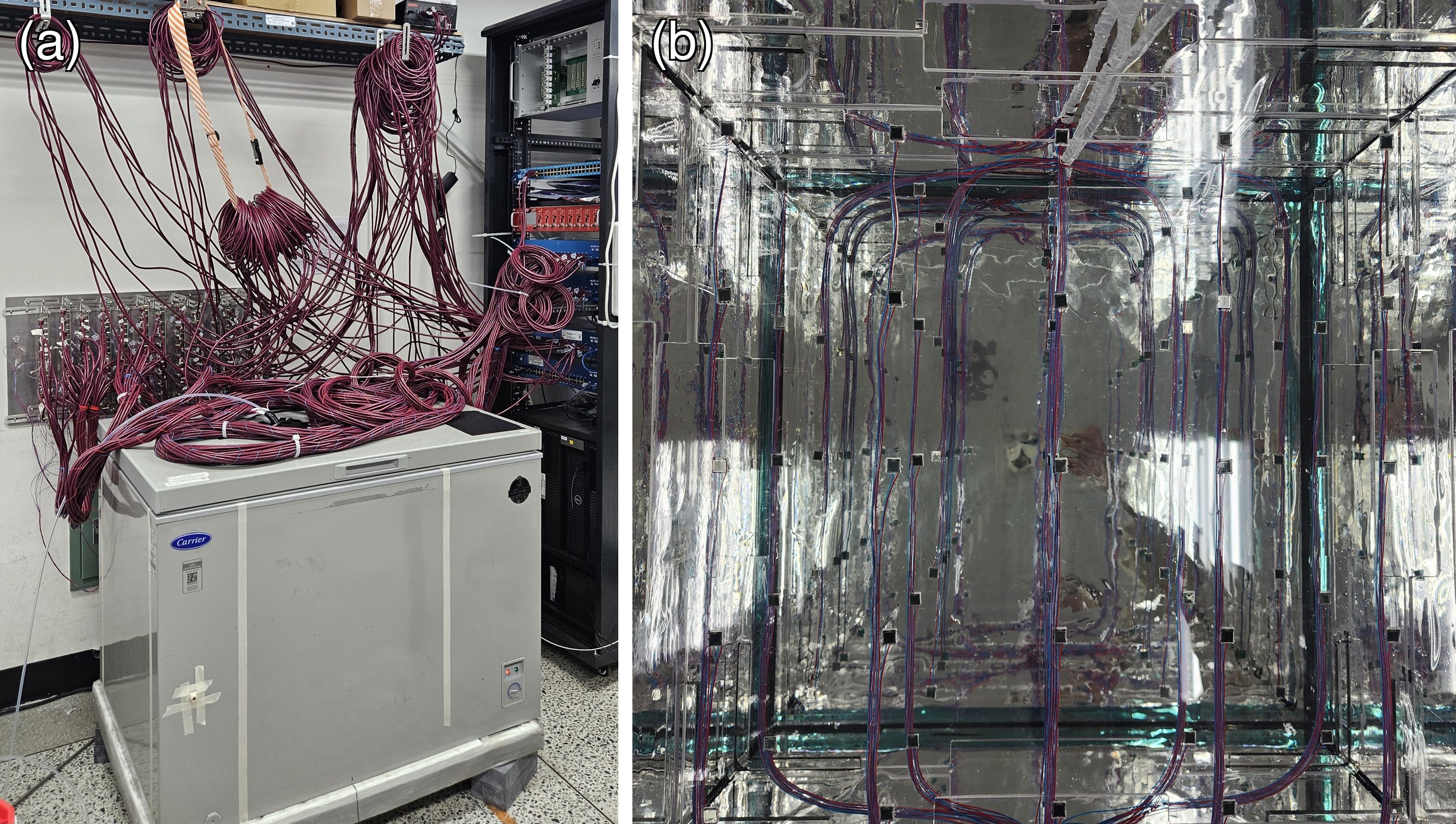}
  \end{center}
  \caption{The CANDY detector. (a) The detector inside the laboratory refrigerator, with the signal and bias cables routed out of the enclosure to the front-end electronics and the DAQ rack. (b) Interior of the acrylic vessel before filling: the $5\times5\times5$ lattice of 125 SiPMs mounted on acrylic plates, with the inner surfaces lined with ESR reflector film.}
  \label{fig:det}
\end{figure*}

In this work, we constructed the CANDY detector, a 125-liter linear alkylbenzene (LAB)-based liquid scintillator detector read out by 125 immersed SiPMs. We characterize its temperature-dependent response from room temperature down to $-30\,^{\circ}\mathrm{C}$. This intermediate operating point was chosen to suppress the sensor dark-count rate while keeping the scintillator in its liquid phase. Section~\ref{sec:setup} describes the detector and the experimental setup. Section~\ref{sec:sipm} describes the SiPM response, including dark counts, SPE gain, and noise occupancy. Section~\ref{sec:scint} describes the scintillation response measured with cosmic-ray muons. Section~\ref{sec:muon} presents the stopping-muon selection with delayed Michel-electron signals and the muon-lifetime measurement at $-30\,^{\circ}\mathrm{C}$. Section~\ref{sec:discussion} discusses the combined implications for future detectors.

\section{Detector and experimental setup}
\label{sec:setup}
\subsection{CANDY detector}

The detector is a cubic acrylic vessel with an active volume of $50\times 50\times 50\,\mathrm{cm}^3$.
Five internal acrylic sheets, spaced $10\,\mathrm{cm}$ apart along the vertical axis,
hold 125 SiPMs arranged in a regular $5\times5\times5$ lattice.
Since the SiPMs are immersed in the scintillator, photons are detected close to their production point with minimal optical transport.
The inner surfaces of the vessel are lined with enhanced specular reflector (ESR) film~\cite{tyvek} to improve photon-collection efficiency and uniformity. Each channel is a Hamamatsu S13360-6075PE SiPM with an active area of $6\times 6\,\mathrm{mm}^2$ and a $75\,\mu\mathrm{m}$ pixel pitch (6400 pixels), a typical breakdown voltage of $53\,\mathrm{V}$ at $25\,^{\circ}\mathrm{C}$ with a temperature coefficient of $54\,\mathrm{mV}/^{\circ}\mathrm{C}$, and a typical gain of $4.0\times10^{6}$ at the recommended $3\,\mathrm{V}$ overvoltage~\cite{sipm}. All 125 channels were operated at a common fixed bias of $56.5\,\mathrm{V}$ throughout the measurement campaign. As the temperature decreases, the breakdown voltage falls, increasing the overvoltage and consequently the gain (Section~\ref{sec:sipm}).

The detector is filled with an LAB-based liquid scintillator containing
$3\,\mathrm{g/L}$ of 2,5-diphenyloxazole (PPO) as the primary fluor and $30\,\mathrm{mg/L}$ of 1,4-bis(2-methylstyryl)benzene (bis-MSB)
as a wavelength shifter~\cite{RENO-LS,Kim:2024spf}.
At $438.5\,\mathrm{nm}$, the refractive indices of acrylic and an LAB-based liquid scintillator with this formulation have been reported as 1.499 and 1.497, respectively~\cite{Yeo2010}.
This close optical match reduces reflection losses at the acrylic--scintillator interfaces.
The same LAB/PPO/bis-MSB scintillator system is used in the COSINE-100 liquid-scintillator veto~\cite{Adhikari:2017esn} and the COSINE-100 Upgrade (COSINE-100U)~\cite{Lee:2024wzd,Park:2026rog,Kim:2024spf}.

To enable low-temperature operation, the detector vessel was sized to fit inside the laboratory refrigerator,
while the front-end electronics remained outside.
Each SiPM was connected to the front-end board through a three-meter-long pair of signal and bias-voltage cables
directly soldered to the sensor.

The detector setup, including the cooling enclosure, is shown in Fig.~\ref{fig:det}.

\subsection{Cooling and temperature control}

\begin{figure}[!htb]
  \begin{center}
      \includegraphics[width=0.45\textwidth]{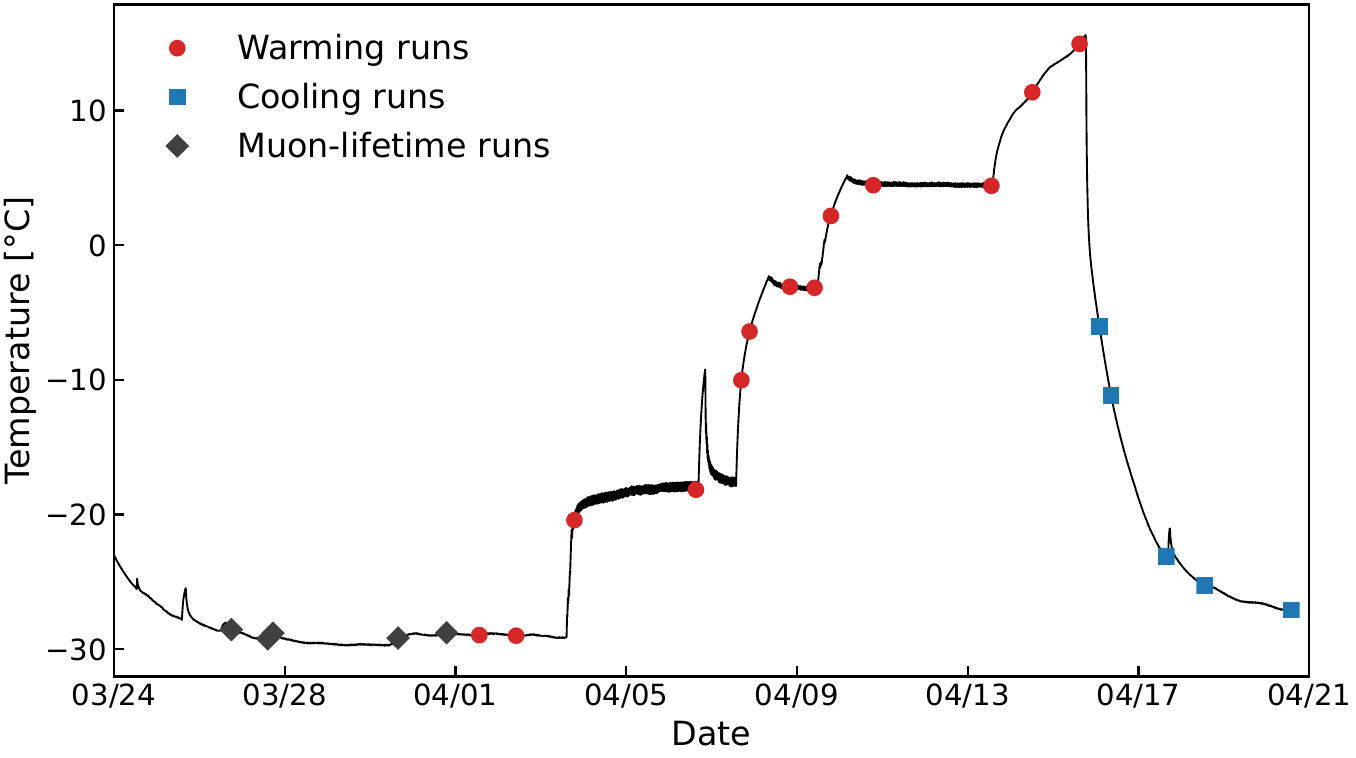}
  \end{center}
  \caption{Temperature measured immediately above the detector vessel during the measurement campaign, logged once per minute. Markers indicate the cosmic-ray muon runs of the temperature scan, taken during the stepped warming (red circles) and the subsequent cooling (blue squares). Diamonds mark five additional runs at $-30\,^{\circ}\mathrm{C}$, recorded before the scan and used for the muon-lifetime measurement of Section~\ref{sec:muon}. The discrete refrigerator settings produce the visible plateaus.}
  \label{fig:temphistory}
\end{figure}

The detector was operated between room temperature and $-30\,^{\circ}\mathrm{C}$ inside a temperature-controlled laboratory refrigerator.
The detector vessel was enclosed in a thermally insulated chamber. A temperature sensor installed inside the refrigerator immediately above the detector vessel monitored the local temperature, which was logged once per minute throughout the measurements.
Figure~\ref{fig:temphistory} shows the recorded temperature history of the measurement campaign: the initial cool-down, an extended plateau near $-30\,^{\circ}\mathrm{C}$, a stepped warming scan up to $+15\,^{\circ}\mathrm{C}$, and a subsequent cooling scan, with the cosmic-ray muon runs marked on the curve.
The refrigerator temperature is adjusted in coarse discrete steps rather than through a continuously tunable setpoint.
At each setting the monitored temperature settled onto a plateau with a root-mean-square (RMS) variation below $0.5\,^{\circ}\mathrm{C}$
and peak-to-peak excursions of about $2\,^{\circ}\mathrm{C}$ from the compressor duty cycle.
To suppress moisture condensation during low-temperature operation, boil-off nitrogen gas was continuously introduced
into the enclosure at a flow rate of approximately $1\,\mathrm{L/min}$.
Detector stability was verified throughout repeated cooling and warming cycles by monitoring waveform baselines and
the response to cosmic-ray muons. No measurable degradation in detector performance was observed over the course of the study.

\FloatBarrier
\begin{figure*}[!b]
  \centering
  \begin{minipage}[t]{0.48\textwidth}
    \vspace{0pt}
    \centering
    \includegraphics[width=\linewidth]{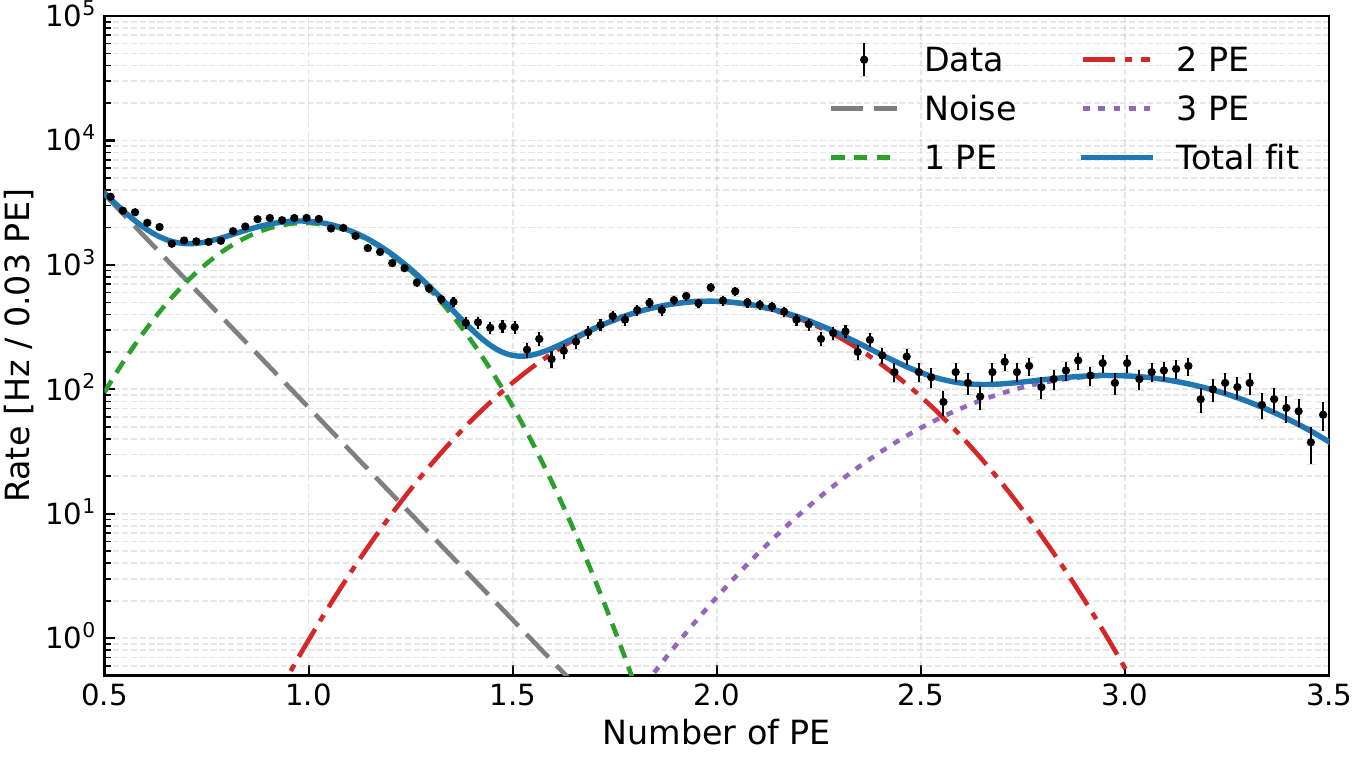}
    \captionof{figure}{Representative pre-trigger dark-count charge spectrum for a single channel at $-30\,^{\circ}\mathrm{C}$, showing the electronics-noise component and resolved $1\,\mathrm{PE}$, $2\,\mathrm{PE}$, and $3\,\mathrm{PE}$ dark-count peaks. Data are shown as points with Poisson uncertainties, overlaid with the Gaussian-fitted total and individual components.}
    \label{fig:darkspectra}
  \end{minipage}\hfill
  \begin{minipage}[t]{0.48\textwidth}
    \vspace{0pt}
    \centering
    \includegraphics[width=\linewidth]{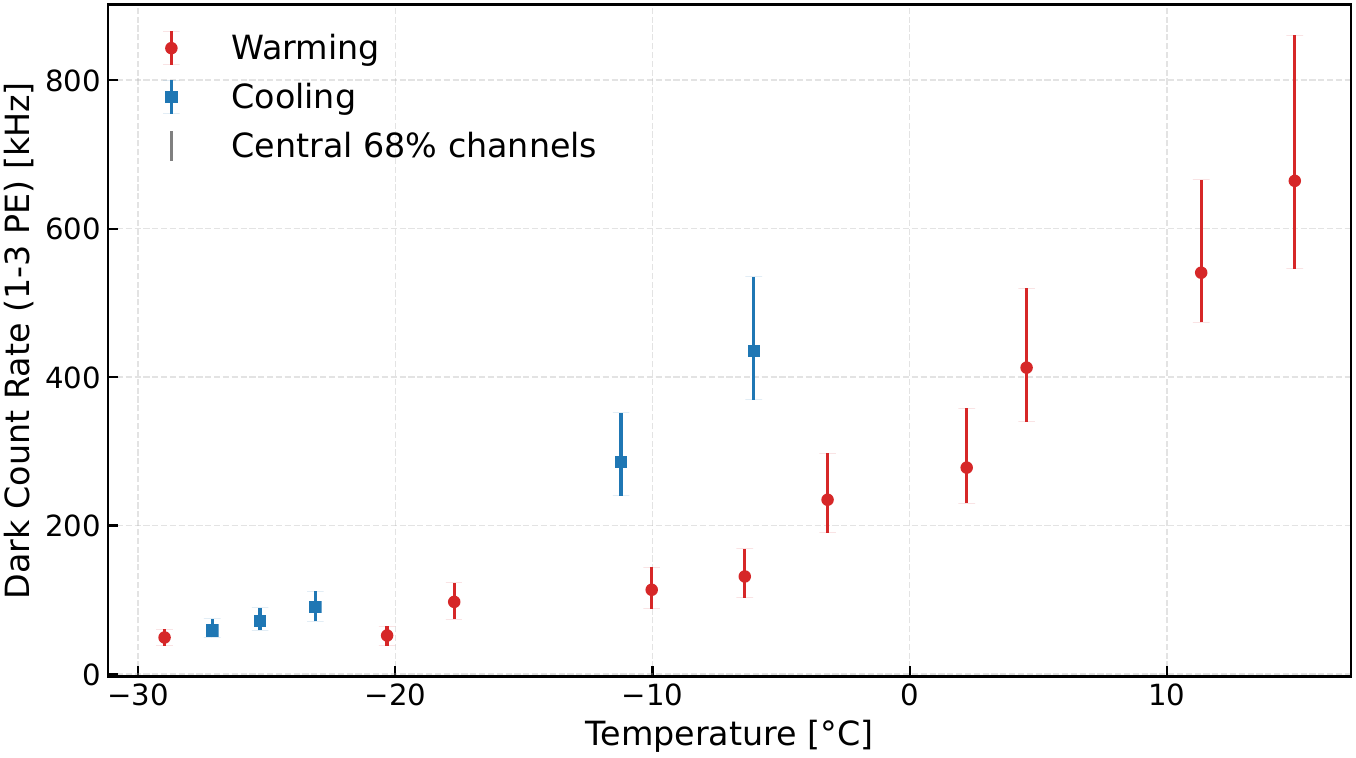}
    \captionof{figure}{Combined $1\,\mathrm{PE}+2\,\mathrm{PE}+3\,\mathrm{PE}$ dark-count rate as a function of detector temperature. Points are medians of channel values, pooling repeated runs at similar temperatures. Error bars show the 16th--84th percentile spread. The rate decreases by about an order of magnitude as the temperature is lowered from $+15\,^{\circ}\mathrm{C}$ to $-30\,^{\circ}\mathrm{C}$.}
    \label{fig:darkrate}
  \end{minipage}
\end{figure*}

\subsection{Readout and data acquisition}

Signal waveforms from all 125 channels were digitized separately by custom waveform digitizers with a $16\,\mathrm{ns}$ sampling interval. An event trigger was issued when at least 5 channels exceeded a per-channel threshold within a $128\,\mathrm{ns}$ coincidence window. The threshold was adjusted with temperature between 175 and 300 analog-to-digital converter (ADC) counts so as to remain at approximately 2 photoelectrons (PE) ($1.9$--$2.8\,\mathrm{PE}$ across runs) despite the changing SiPM gain.
For each trigger, full waveforms from all channels were recorded over a $4\,\mu\mathrm{s}$ acquisition window,
with the trigger positioned approximately $1.2\,\mu\mathrm{s}$ after the start of the waveform.
The $480\,\mathrm{ns}$ integration interval centered on the trigger is referred to as the main-signal window.
The recorded waveforms were analyzed channel by channel using channel-specific pedestal and SPE calibrations. The resulting calibrated quantities were combined only afterward, where required, to construct detector-level observables.
The pre-trigger window was used to measure the SiPM dark-count rate.

The dataset consists of self-triggered events dominated by cosmic-ray muons. Temperature scans were performed during both warm-up and cool-down cycles while continuously recording data, with each temperature point containing approximately $3\times10^5$ triggered events. Five additional runs were recorded at $-30\,^{\circ}\mathrm{C}$ before the scan and are used, together with the coldest scan runs, for the muon-lifetime measurement of Section~\ref{sec:muon} (Fig.~\ref{fig:temphistory}). In total, about $6.5\times10^{6}$ events were recorded over a live time of $40{,}807\,\mathrm{s}$.

\section{Temperature-dependent SiPM response}
\label{sec:sipm}

We characterize the temperature-dependent SiPM response using two regions of the recorded waveforms.
Dark-count rates are measured in the first $800\,\mathrm{ns}$, before the trigger, whereas the SPE response is extracted from charge integrated over the $960$--$1440\,\mathrm{ns}$ main-signal window. Both intervals are measured from the start of the waveform.
The measurements are performed channel by channel at each temperature.
The fitted SPE charge separation defines the channel-specific ADC-to-PE conversion used throughout this work.

\subsection{Dark-count rate}

The pre-trigger dark-count charge spectrum shows well-separated single-photoelectron ($1\,\mathrm{PE}$), $2\,\mathrm{PE}$, and $3\,\mathrm{PE}$ peaks on top of the electronics noise, as shown in Fig.~\ref{fig:darkspectra}. The rate of these dark-count peaks depends strongly on temperature.

Figure~\ref{fig:darkrate} summarizes the temperature dependence of the combined $1\,\mathrm{PE}+2\,\mathrm{PE}+3\,\mathrm{PE}$ dark-count rate. As the temperature decreased from $+15\,^{\circ}\mathrm{C}$ to $-30\,^{\circ}\mathrm{C}$, the per-channel median dark-count rate reduced from $664.3\,\mathrm{kHz}$ to $49.2\,\mathrm{kHz}$, a factor of 13.5, consistent with thermally activated carrier generation in the SiPMs~\cite{Acerbi2019}.

The reduced dark-count rate at low temperature directly lowers the noise occupancy of each channel, defined as the probability that a channel registers at least one dark count above the $\sim2\,\mathrm{PE}$ per-channel trigger threshold within the $128\,\mathrm{ns}$ trigger coincidence window, $p = 1-e^{-r\,\Delta t}$ for the corresponding $2$--$3\,\mathrm{PE}$ dark-count rate $r$. The per-channel median occupancy falls from about 2.12\% at $+15\,^{\circ}\mathrm{C}$ to about 0.18\% at $-30\,^{\circ}\mathrm{C}$, strongly suppressing the accidental contribution to the trigger.

\begin{figure*}[!t]
  \centering
  \begin{minipage}[t]{0.48\textwidth}
    \vspace{0pt}
    \centering
    \includegraphics[width=\linewidth]{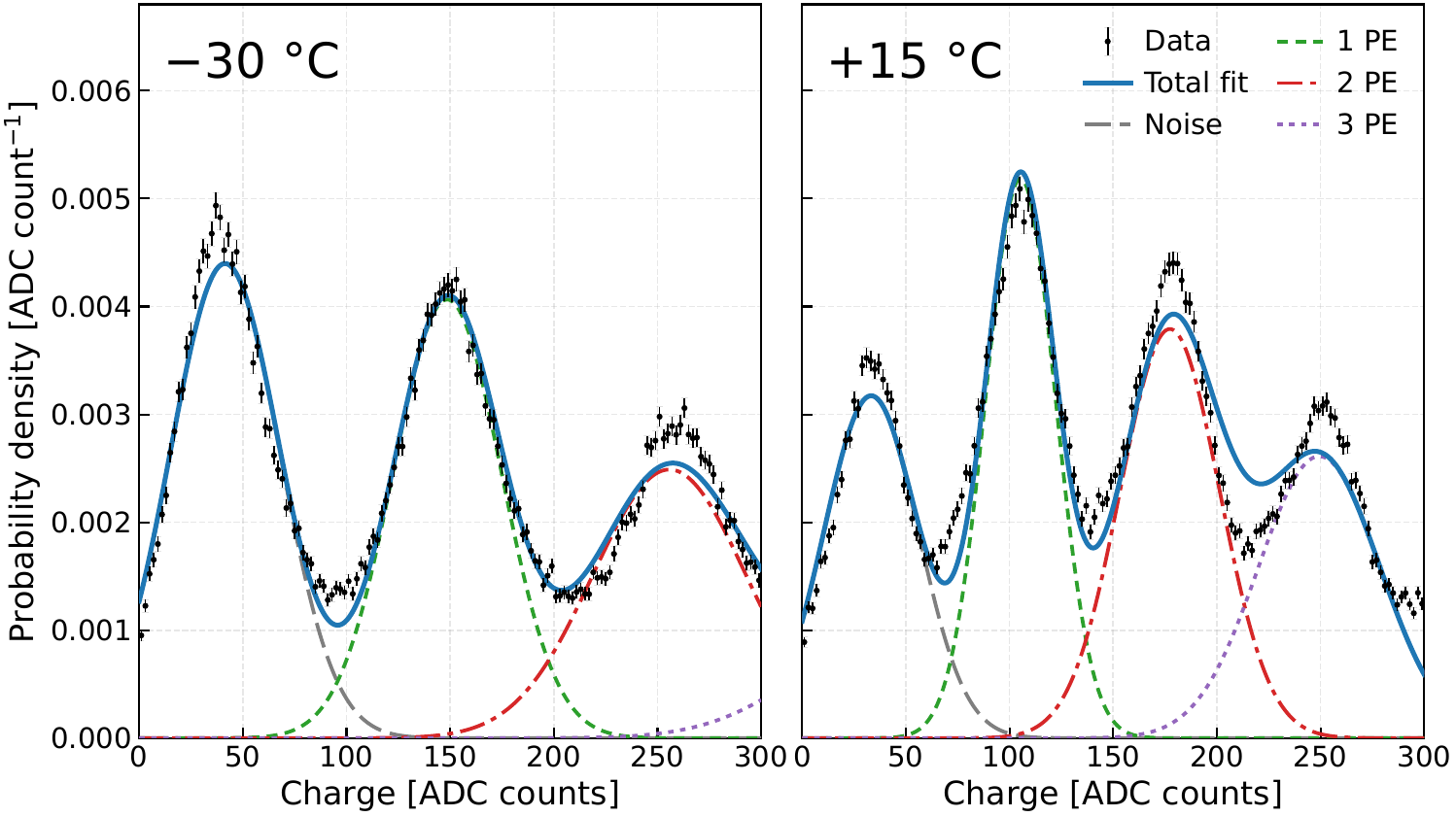}
    \captionof{figure}{Main-signal-window charge spectra for a representative channel at low ($-30\,^{\circ}\mathrm{C}$) and high ($+15\,^{\circ}\mathrm{C}$) temperature, with the fitted noise, $1\,\mathrm{PE}$, $2\,\mathrm{PE}$, and $3\,\mathrm{PE}$ components. The improved pedestal-to-$1\,\mathrm{PE}$ separation at low temperature yields a more stable gain calibration.}
    \label{fig:speexamples}
  \end{minipage}\hfill
  \begin{minipage}[t]{0.48\textwidth}
    \vspace{0pt}
    \centering
    \includegraphics[width=\linewidth]{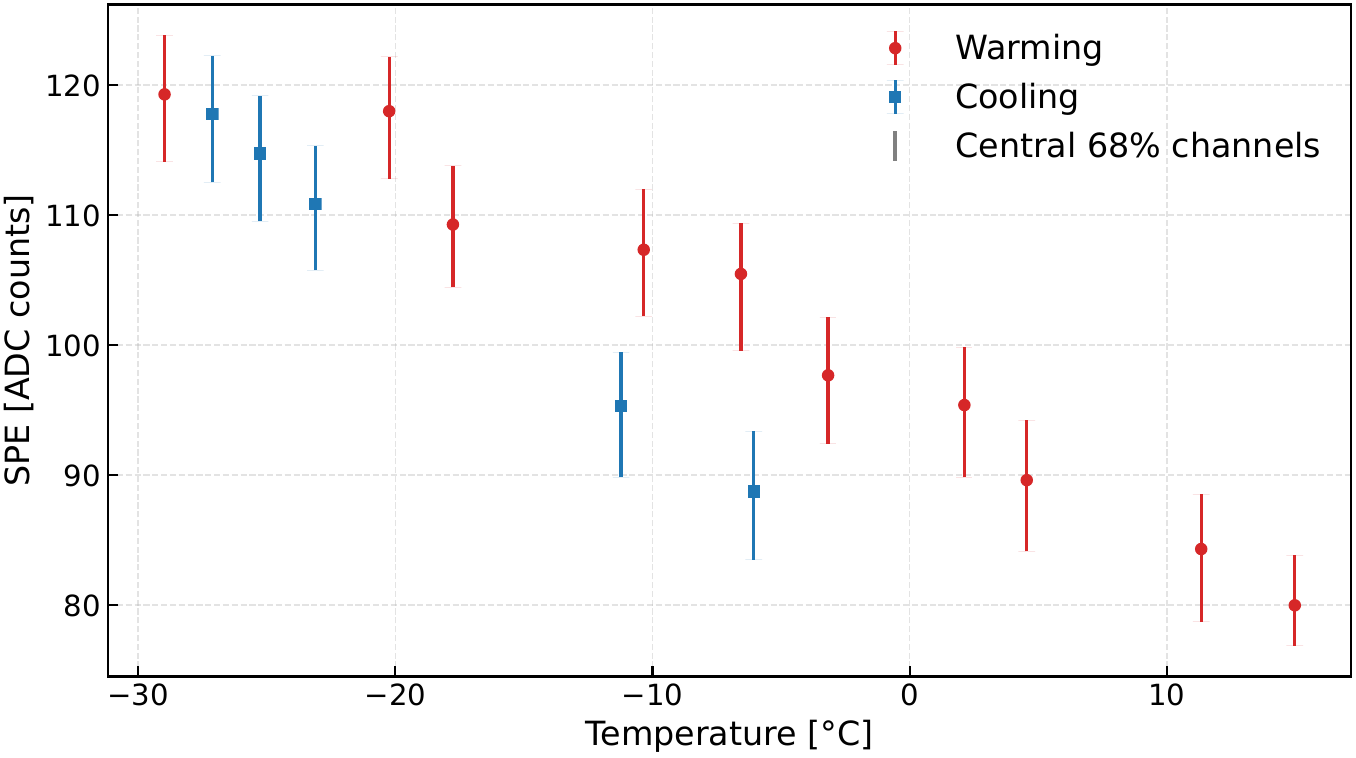}
    \captionof{figure}{Single-photoelectron peak position (proportional to SiPM gain) as a function of detector temperature, for both warming and cooling runs. Points are medians of channel values, pooling repeated runs at similar temperatures. Error bars show the 16th--84th percentile spread. The response increases by 49\% as the temperature is lowered from $+15\,^{\circ}\mathrm{C}$ to $-30\,^{\circ}\mathrm{C}$.}
    \label{fig:spevsT}
  \end{minipage}
\end{figure*}

\subsection{Single-photoelectron response}

The SPE response was extracted from the per-channel main-signal-window charge distributions.
Figure~\ref{fig:speexamples} compares main-signal-window charge spectra measured at low and high temperature for a representative channel.
The separation between the pedestal and the single-photoelectron peak becomes clearer at low temperature. This is quantified by the separation figure of merit
\[
\mathrm{FoM} = \frac{\left|\mu_{\mathrm{ped}} - \mu_{1\mathrm{PE}}\right|}{\sqrt{\sigma_{\mathrm{ped}}^{2} + \sigma_{1\mathrm{PE}}^{2}}},
\]
where $\mu$ and $\sigma$ denote the Gaussian-fitted positions and widths of the pedestal and $1\,\mathrm{PE}$ peaks, respectively. The per-channel median FoM improves from 4.3 at $+15\,^{\circ}\mathrm{C}$ to 5.5 at $-30\,^{\circ}\mathrm{C}$.

The fitted SPE charge separation, used as a proxy for the SiPM gain, had a per-channel median of $80.0$ ADC counts at $+15\,^{\circ}\mathrm{C}$ and $119.3$ ADC counts at $-30\,^{\circ}\mathrm{C}$ (Fig.~\ref{fig:spevsT}). Their ratio, $119.3/80.0=1.49$, corresponds to a 49\% increase in gain upon cooling. At the fixed $56.5\,\mathrm{V}$ bias, the $45\,^{\circ}\mathrm{C}$ temperature span corresponds to a breakdown-voltage decrease, and hence an overvoltage increase, of approximately $2.4\,\mathrm{V}$ based on the specified temperature coefficient of $54\,\mathrm{mV}/^{\circ}\mathrm{C}$~\cite{Acerbi2019,sipm}. Assuming that the gain is proportional to the overvoltage, the relation $1.49=(V_{\mathrm{ov,warm}}+2.4\,\mathrm{V})/V_{\mathrm{ov,warm}}$ gives inferred overvoltages of approximately $4.9\,\mathrm{V}$ at $+15\,^{\circ}\mathrm{C}$ and $7.3\,\mathrm{V}$ at $-30\,^{\circ}\mathrm{C}$. These values are consistent with the typical breakdown voltage of the device within its specified tolerance.

\FloatBarrier
\section{Temperature-dependent scintillation response}
\label{sec:scint}

The detector-level scintillation response was measured using cosmic-ray muons traversing the active volume.

\subsection{Cosmic-ray muon spectra and rates}

\begin{figure*}[!htb]
  \begin{center}
      \includegraphics[width=0.85\textwidth]{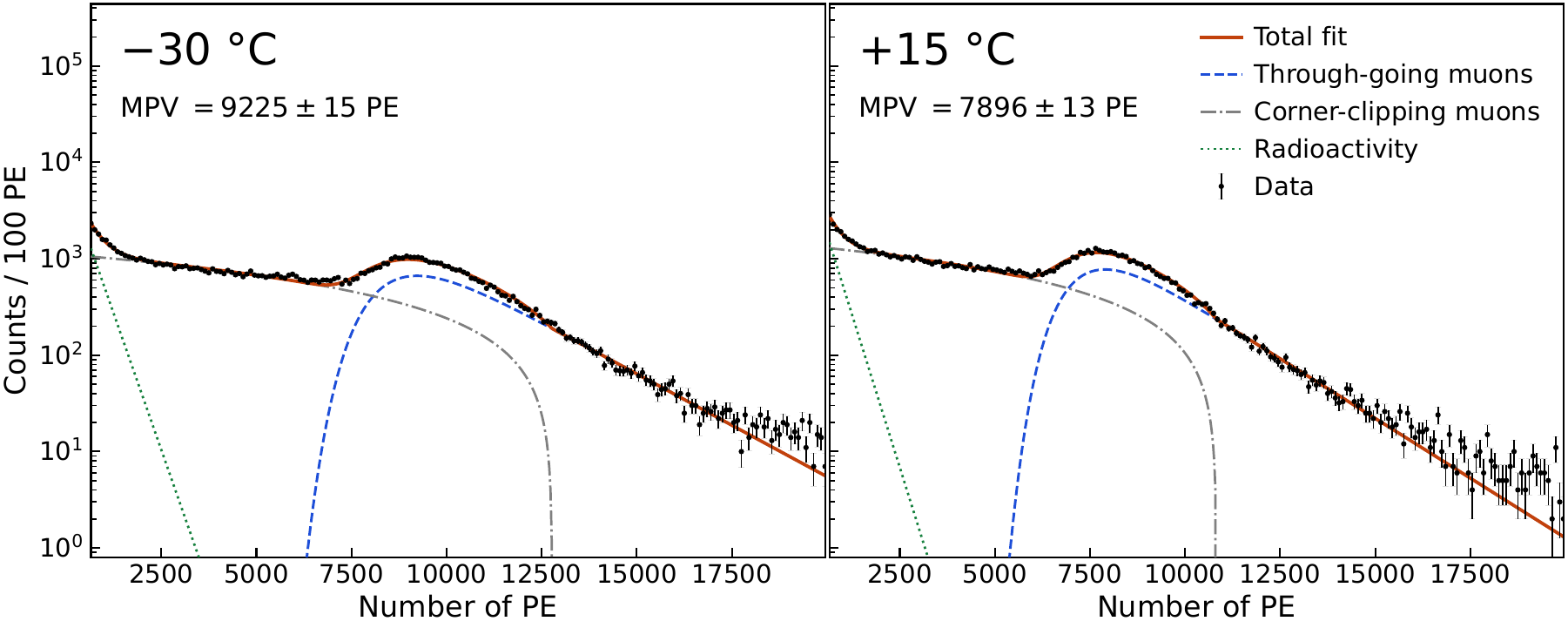}
  \end{center}
  \caption{Total photon-count spectra of cosmic-ray muons at low ($-30\,^{\circ}\mathrm{C}$, left) and high ($+15\,^{\circ}\mathrm{C}$, right) temperature, decomposed into through-going muons (Landau), corner-clipping muons (linear with cutoff), and environmental radioactivity (exponential). The Landau MPV increases from $7896\pm13\,\mathrm{PE}$ at $+15\,^{\circ}\mathrm{C}$ to $9225\pm15\,\mathrm{PE}$ at $-30\,^{\circ}\mathrm{C}$, while the through-going muon rate obtained from the Landau integral is consistent at both temperatures.}
  \label{fig:muonspectra}
\end{figure*}

Figure~\ref{fig:muonspectra} shows the total photon-count spectra at the coldest ($-30\,^{\circ}\mathrm{C}$) and warmest ($+15\,^{\circ}\mathrm{C}$) points. Over the range $600$--$20{,}000\,\mathrm{PE}$, each spectrum is fitted with the sum of three components: a Landau distribution for muons traversing the full detector height (through-going muons), a geometrically motivated linear term with a cutoff for muons that clip a corner of the volume with a partial path length (corner-clipping muons), and a steep exponential for environmental radioactivity at low charge. The fit describes the data over the full range, and the Landau most probable value (MPV) increases from $7896\pm13\,\mathrm{PE}$ at $+15\,^{\circ}\mathrm{C}$ to $9225\pm15\,\mathrm{PE}$ at $-30\,^{\circ}\mathrm{C}$, corresponding to a 16.8\% increase in the detected light yield for these individual runs after the per-channel SPE calibration.

\begin{figure}[!htb]
  \begin{center}
      \includegraphics[width=0.45\textwidth]{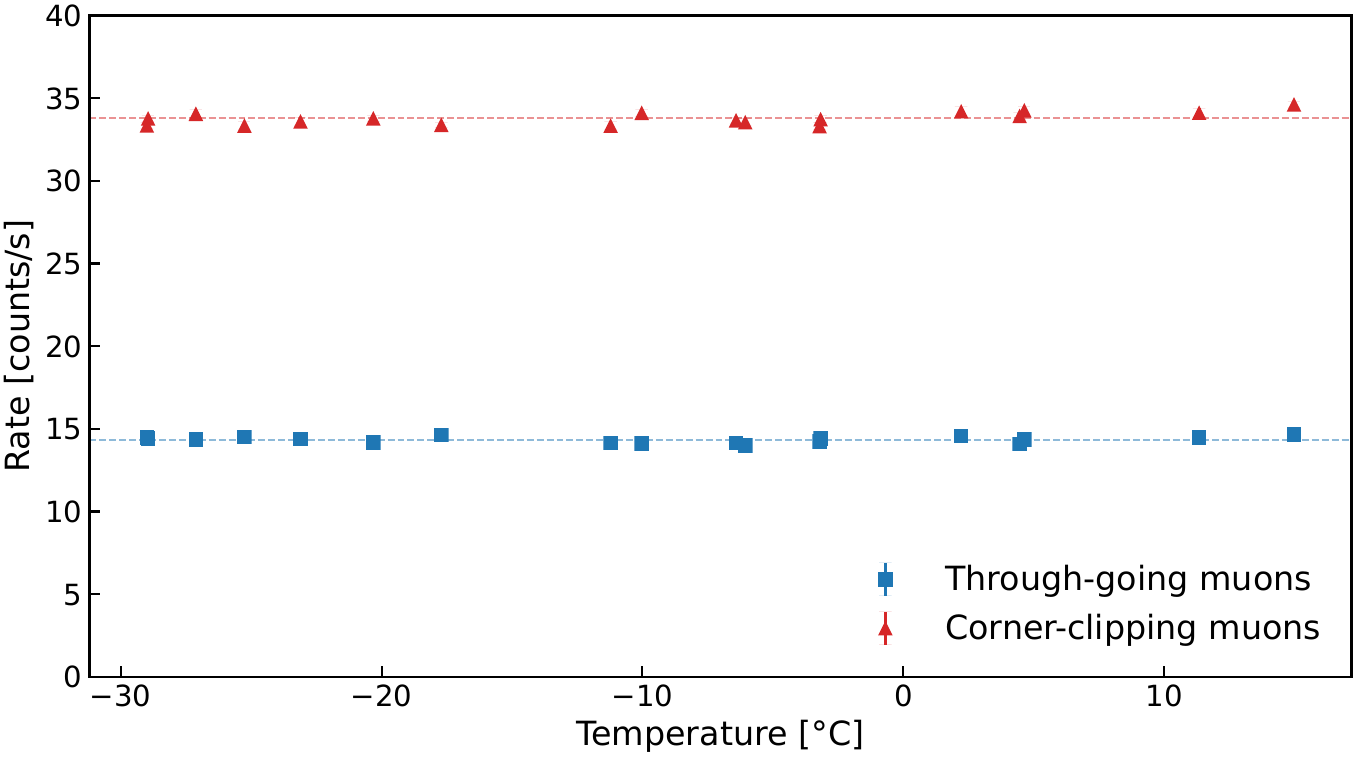}
  \end{center}
  \caption{Detected rates of through-going and corner-clipping muons as a function of temperature, obtained from the component integrals of the spectral decomposition of Fig.~\ref{fig:muonspectra}. Error bars are propagated fit-parameter uncertainties, mostly smaller than the markers, and the dashed lines mark the mean rates. Both rates are constant over the full temperature range.}
  \label{fig:muonrate}
\end{figure}

The analytic integrals of the Landau and linear components, divided by the run live time, give the detected rates of through-going and corner-clipping muons. Figure~\ref{fig:muonrate} shows both rates for all 18 runs of the temperature scan: the through-going rate is $14.3\pm0.2\,\mathrm{s}^{-1}$ and the corner-clipping rate is $33.8\pm0.4\,\mathrm{s}^{-1}$, for a total muon rate of $48.1\pm0.5\,\mathrm{s}^{-1}$, all constant over the full temperature range.

\subsection{Light-yield variation}

\begin{figure}[!htb]
  \begin{center}
      \includegraphics[width=0.45\textwidth]{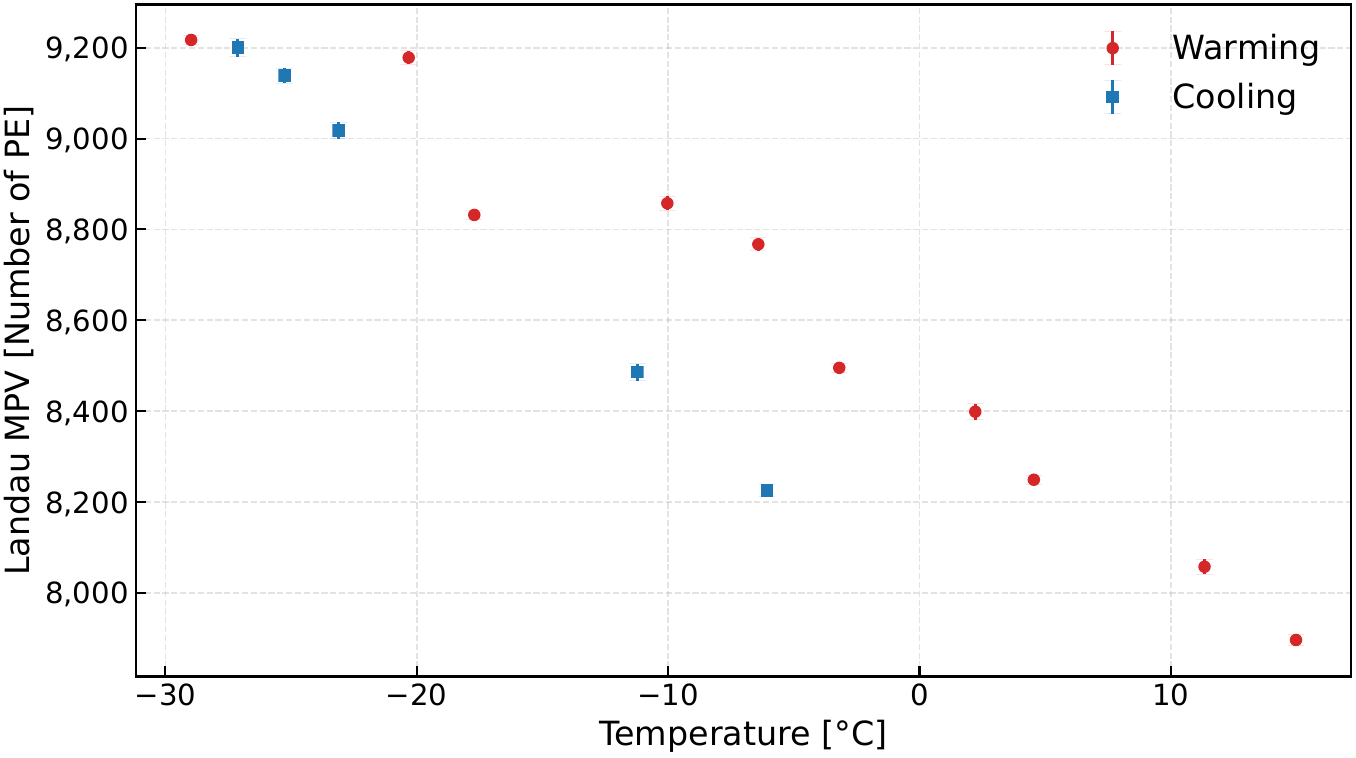}
  \end{center}
  \caption{Cosmic-ray muon Landau MPV (detector light yield) as a function of temperature for warming and cooling runs, from the spectral decomposition of Fig.~\ref{fig:muonspectra}. Error bars are the MPV fit-parameter uncertainties. The light yield increases by 17\% as the temperature is lowered from $+15\,^{\circ}\mathrm{C}$ to $-30\,^{\circ}\mathrm{C}$.}
  \label{fig:landauvsT}
\end{figure}

The detector light yield was tracked using the Landau MPV in PE from the spectral decomposition described above, applied to each temperature point of the scan.
Figure~\ref{fig:landauvsT} shows its evolution with temperature for warming and cooling runs separately, with uncertainties given by the MPV fit-parameter errors. Pooling the two $-30\,^{\circ}\mathrm{C}$ warming-scan runs gives a 16.7\% increase in light yield relative to the $+15\,^{\circ}\mathrm{C}$ run, while the SPE gain increased by 49.1\%. Their product, $1.167\times1.491\simeq1.74$, is consistent with the factor of 1.74 increase in the Landau peak in raw ADC units, from about $6.3\times10^{5}$ to about $1.1\times10^{6}$ ADC counts, before the per-channel SPE correction.

\subsection{Waveform evolution}

\begin{figure*}[!htb]
  \begin{center}
      \includegraphics[width=0.85\textwidth]{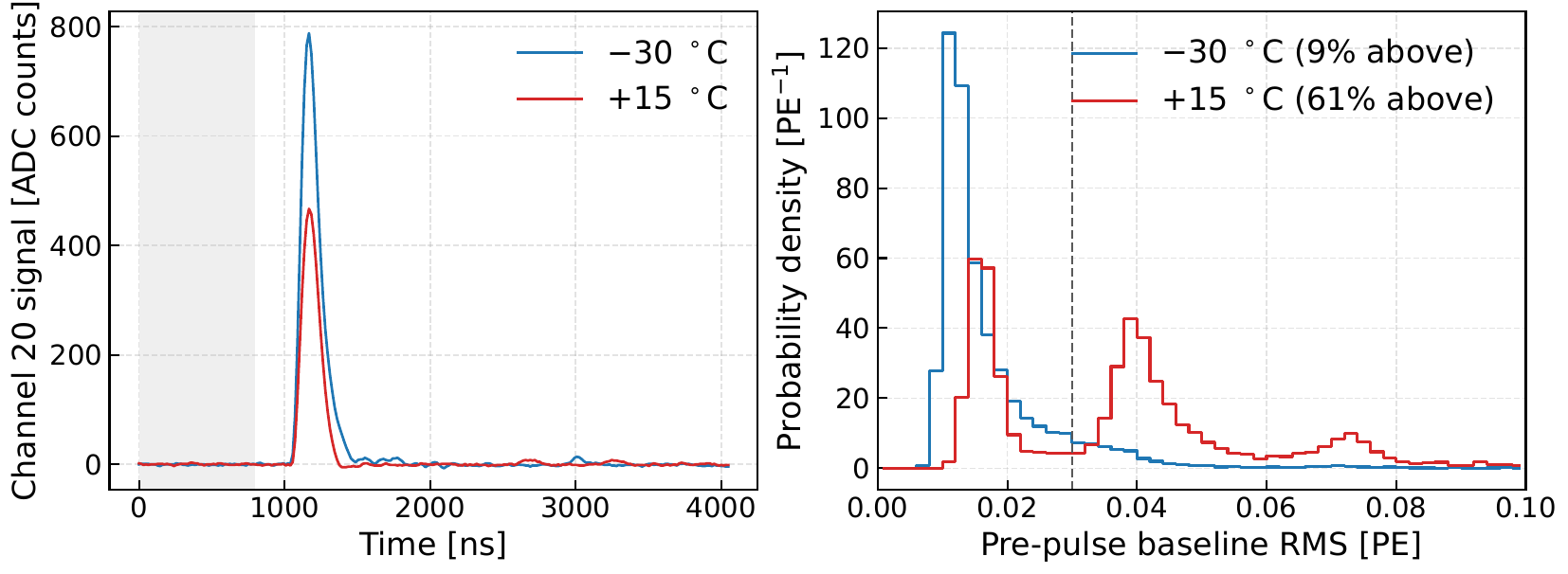}
  \end{center}
  \caption{Left: pedestal-subtracted raw-ADC waveforms of channel 20 at low ($-30\,^{\circ}\mathrm{C}$) and high ($+15\,^{\circ}\mathrm{C}$) temperature. The shaded band marks the pre-pulse region. At each temperature, events within 5\% of the fitted Landau MPV were aligned using the leading edge of the all-channel summed waveform to construct the mean waveform for channel 20. The displayed trace is the actual event within 1\% of the MPV that is closest to the corresponding mean. Right: event-by-event distribution of the pre-pulse baseline RMS of channel 20 in SPE-normalized units. The dashed line at $0.03\,\mathrm{PE}$ marks the minimum separating the dark-count-free core from the baselines holding a dark count. At $+15\,^{\circ}\mathrm{C}$, 61\% of the events lie above it, compared with 9\% at $-30\,^{\circ}\mathrm{C}$.}
  \label{fig:waveformcomparison}
\end{figure*}

The reduced dark-count activity at low temperature also produces cleaner waveform baselines. Figure~\ref{fig:waveformcomparison} shows representative raw-ADC waveforms of channel 20 selected from cosmic-ray muon events near the fitted Landau MPV at low and high temperature (left) and the event-by-event distribution of its pre-pulse baseline RMS (right). The larger pulse amplitude at low temperature is consistent with the independently measured increases in the SPE gain and light yield discussed above. At $+15\,^{\circ}\mathrm{C}$ the RMS distribution is bimodal: a dark-count-free core near $0.016\,\mathrm{PE}$ and a second population near $0.040\,\mathrm{PE}$ from baselines in which a dark count landed in the pre-pulse window. Counting the events above $0.03\,\mathrm{PE}$, the minimum separating the two, 61\% of the baselines hold a dark count at $+15\,^{\circ}\mathrm{C}$ against 9\% at $-30\,^{\circ}\mathrm{C}$. The clean cores themselves also move apart, from $0.016\,\mathrm{PE}$ to $0.011\,\mathrm{PE}$, as the larger gain reduces the fixed electronics noise relative to one photoelectron.

\section{Stopping-muon and Michel-electron measurement}
\label{sec:muon}

To measure the muon lifetime at $-30\,^{\circ}\mathrm{C}$, we select events with a prompt muon pulse followed by a delayed Michel-electron pulse. For a positive muon, the decay proceeds as
\[
\mu^+ \rightarrow e^+ + \nu_e + \bar{\nu}_\mu,
\]
with a free-muon lifetime of $\tau_{\mu^+} = 2197\,\mathrm{ns}$~\cite{PDG-mu}.

\subsection{Double-pulse selection}

\begin{figure}[!htb]
  \begin{center}
      \includegraphics[width=0.45\textwidth]{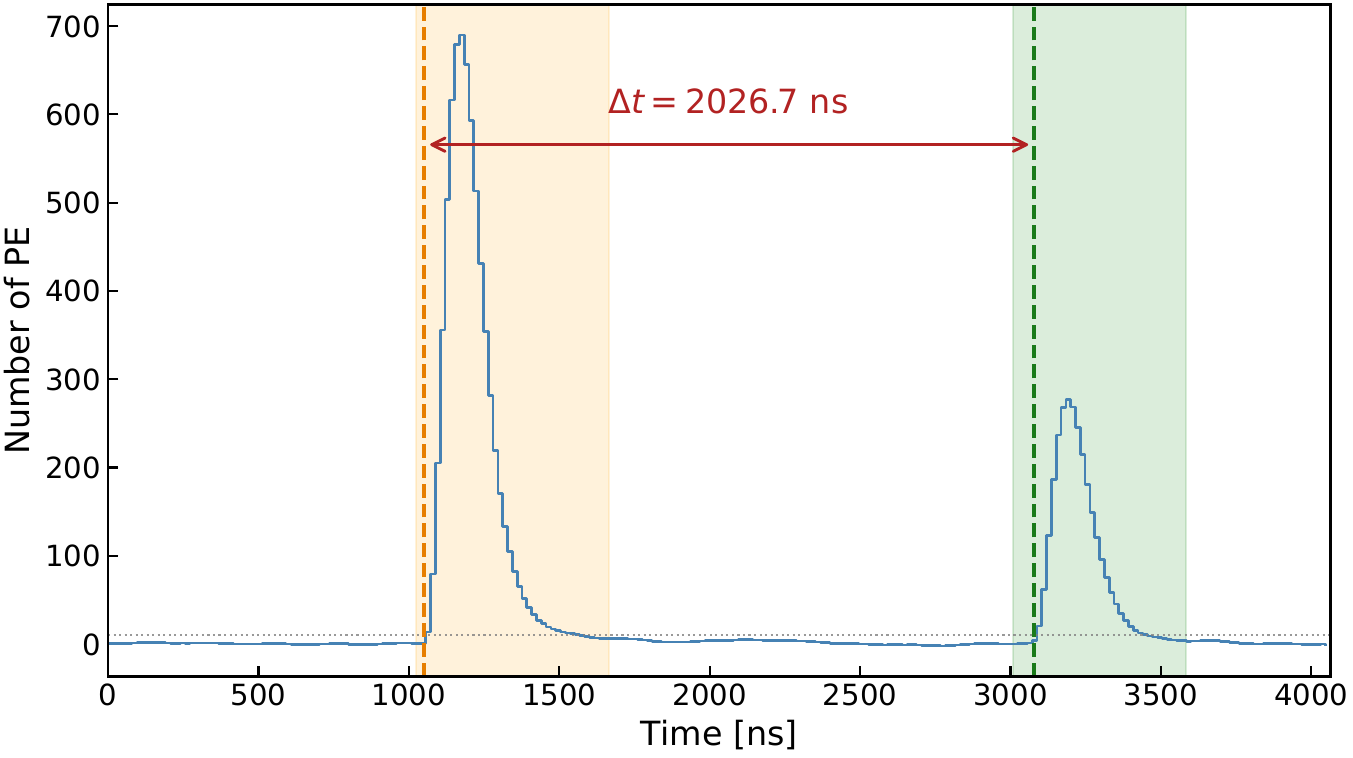}
  \end{center}
  \caption{Example double-pulse waveform of a stopping-muon candidate. The prompt pulse is the stopping muon and the delayed pulse the Michel electron. The cluster windows, leading edges, and time separation $\Delta t$ are indicated.}
  \label{fig:doublepulse}
\end{figure}

The measurement uses the data taken at $-30\,^{\circ}\mathrm{C}$: the five runs recorded before the temperature scan and the two coldest scan runs (Fig.~\ref{fig:temphistory}), for a combined live time of about $10{,}500\,\mathrm{s}$. Stopping-muon candidates were identified from double-pulse signatures in the waveform summed over all channels. The prompt pulse corresponds to the stopping muon and the delayed pulse to the Michel electron. For each trigger, the summed waveform was processed with charge integration ($Q$) and leading-edge timing, and two temporally separated pulses were required within the $4\,\mu\mathrm{s}$ acquisition window. To suppress low-energy backgrounds and noise, the prompt and delayed pulses were required to satisfy $Q_1>1000\,\mathrm{PE}$ and $Q_2>1000\,\mathrm{PE}$, respectively. With this selection, 3{,}470 stopping-muon candidates were retained. An example double-pulse candidate is shown in Fig.~\ref{fig:doublepulse}.

\subsection{Muon lifetime}

\begin{figure}[!htb]
  \begin{center}
      \includegraphics[width=0.45\textwidth]{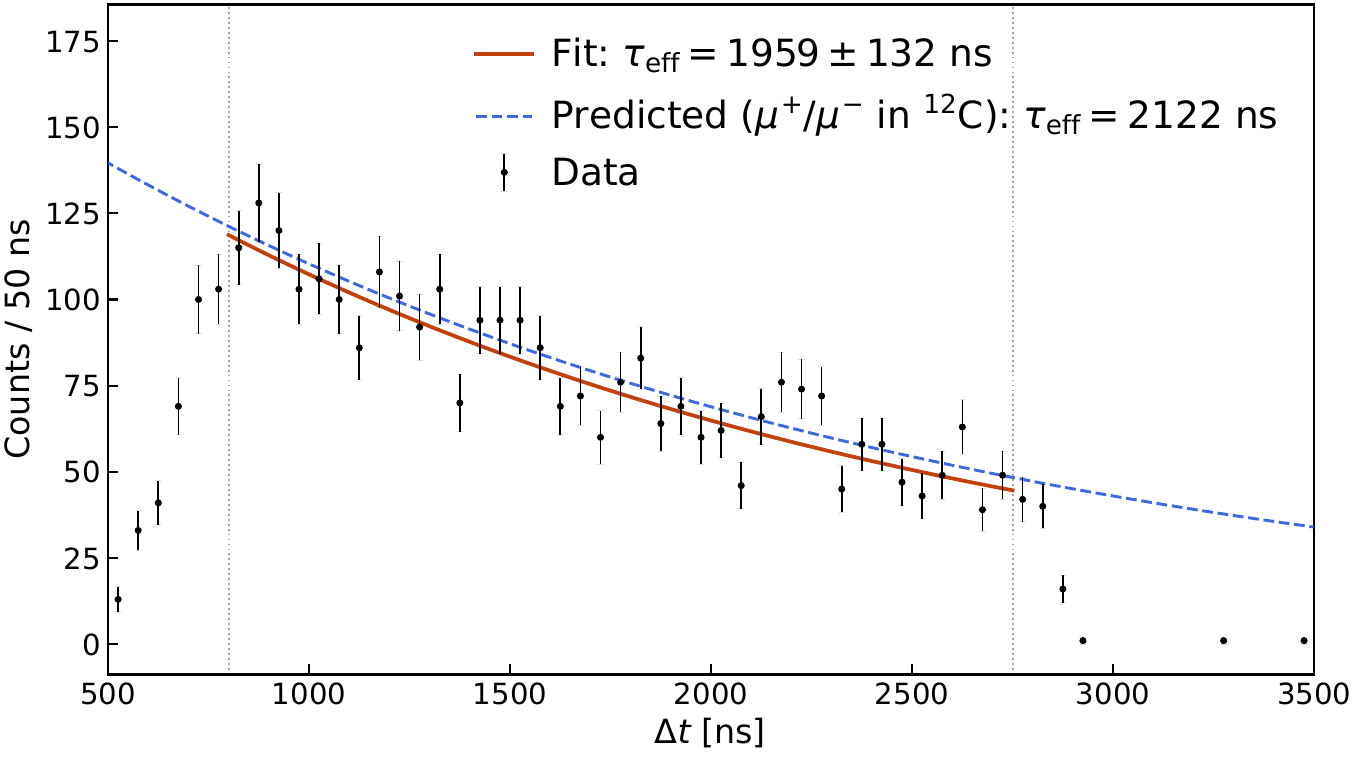}
  \end{center}
  \caption{Distribution of the prompt-to-delayed time difference $\Delta t$ for the 3{,}470 stopping-muon candidates recorded at $-30\,^{\circ}\mathrm{C}$, with the best-fit exponential-plus-constant model (the accidental term $C_\mu$ is fixed at 1.19652 counts per bin, not fitted). The fitted region is $0.8$--$2.75\,\mu\mathrm{s}$ and the fitted effective lifetime is $\tau_{\mathrm{eff}}=1959\pm132\,\mathrm{ns}$, in agreement with the predicted effective curve for a mixed $\mu^{+}/\mu^{-}$ population in $^{12}$C ($\tau_{\mathrm{eff}}\approx2122\,\mathrm{ns}$, dashed).}
  \label{fig:lifetime}
\end{figure}

The distribution of the time difference $\Delta t$ between the prompt and delayed pulses was fitted with an exponential decay plus a constant accidental background,
\[
f(\Delta t)=A\,e^{-\Delta t/\tau}+C_\mu ,
\]
where the constant term $C_\mu$ accounts for accidental coincidences and is fixed rather than fitted. The fit window of $0.8$--$2.75\,\mu\mathrm{s}$ follows from the acquisition structure: with the prompt pulse near $1.2\,\mu\mathrm{s}$ in the $4\,\mu\mathrm{s}$ window, the lower bound keeps the delayed pulse well separated from the prompt pulse, and the upper bound corresponds to the end of the acquisition window. The accidental background is estimated from the $N_{\mathrm{prompt}}=497{,}515$ events passing the prompt-pulse selection, before requiring a second pulse. Using the total muon rate $R_\mu=48.1\,\mathrm{s}^{-1}$ measured in Section~\ref{sec:scint} and a bin width of $\Delta t_{\mathrm{bin}}=50\,\mathrm{ns}$, we obtain $C_\mu=N_{\mathrm{prompt}}R_\mu\Delta t_{\mathrm{bin}}=1.19652$ counts per bin. This corresponds to 46.7 accidental events, or 1.56\% of the 3{,}000 entries in the fitted interval, indicating that the selected sample remains dominated by stopping-muon Michel-electron events. The measured distribution and best-fit curve are shown in Fig.~\ref{fig:lifetime}.

\begin{samepage}
The single-exponential fit returns the population-averaged \emph{effective} lifetime
\[
\tau_{\mathrm{eff}} = 1959 \pm 132\,\mathrm{ns}.
\]
\end{samepage}

This value should not be compared directly with the free muon lifetime, because the LAB scintillator is a hydrocarbon: stopped negative muons do not all decay but a fraction undergo nuclear capture on $^{12}$C, which shortens their disappearance time to $\tau_{\mu^-}\approx2028\,\mathrm{ns}$~\cite{Measday}, whereas positive muons decay with the free lifetime. The recorded $\Delta t$ distribution is therefore a mixture of the two populations, and a single exponential returns an effective lifetime that lies below the free value. Combining the two components with an assumed charge ratio $\mu^+/\mu^- = 1.27$, motivated by the WILLI low-energy measurement~\cite{WILLI1999}, predicts an effective lifetime of about $2122\,\mathrm{ns}$ over the fitted window, which is $1.24\sigma$ above the measured value. As a cross-check, applying the same selection and fit to the five scan runs taken above $0\,^{\circ}\mathrm{C}$ yields $\tau_{\mathrm{eff}} = 2034 \pm 159\,\mathrm{ns}$, consistent with the $-30\,^{\circ}\mathrm{C}$ result and with the expectation.

\section{Discussion}
\label{sec:discussion}

The measurements above show that the three temperature effects act together. Cooling from $+15\,^{\circ}\mathrm{C}$ to $-30\,^{\circ}\mathrm{C}$ suppressed the SiPM dark-count rate by a factor of 13.5, increased the SPE gain by 49.1\%, and increased the detected cosmic-ray muon light yield by 16.7\%. The gain increase and the increase in the SPE-calibrated PE signal are distinct contributions to the larger raw-ADC response. Characterizing both \emph{in situ} for the full immersed array distinguishes changes in charge amplification from changes in the detected light signal.

The warming and cooling scans in Figs.~\ref{fig:spevsT} and~\ref{fig:landauvsT} show a small systematic offset between the two directions. The reported temperature was measured just above the detector vessel, rather than directly in the liquid scintillator. Data were collected during both warming and cooling, without waiting for the liquid scintillator to reach thermal equilibrium. Because of the large heat capacity of the 125-liter volume, the liquid temperature lags the sensor reading. This lag produces temperature offsets of opposite signs during warming and cooling, leading to the observed hysteresis. The two scans therefore bracket the equilibrium response, and their difference can be taken as a conservative estimate of the temperature-scale uncertainty.

The reduced dark-count rate lowers the multi-photoelectron noise occupancy and reduces noise contributions to the trigger multiplicity. Cooling also improves pedestal-to-$1\,\mathrm{PE}$ separation and yields cleaner waveform baselines. The cleaner baselines and improved photoelectron separation help identify the prompt and delayed pulses used in the muon-lifetime measurement.

\section{Conclusion}
\label{sec:conclusion}

We have characterized a 125-liter LAB-based liquid scintillator detector read out by 125 immersed SiPM channels, operated from room temperature to $-30\,^{\circ}\mathrm{C}$. Cooling from $+15\,^{\circ}\mathrm{C}$ to $-30\,^{\circ}\mathrm{C}$ reduced the SiPM dark-count rate by a factor of 13.5, increased the SPE response by 49.1\%, and increased the cosmic-ray muon light yield by 16.7\%. The improved signal-to-noise ratio at low temperature enabled a clean stopping-muon Michel-electron selection at $-30\,^{\circ}\mathrm{C}$, from which the effective muon lifetime was measured to be $1959\pm132\,\mathrm{ns}$, consistent with the value expected for a hydrocarbon scintillator after accounting for $\mu^{-}$ capture on carbon. These results demonstrate the feasibility and stability of immersed SiPM arrays in liquid scintillator at sub-zero temperatures and provide reference data for future distributed-photosensor scintillation detectors.

\section*{Acknowledgments}

This research was supported by the Chung-Ang University Graduate Research Scholarship in 2025
and by the National Research Foundation of Korea (NRF) grant funded
by the Korean government (MSIT) (RS-2021-NR058750 and RS-2024-00438814).

\bibliography{candyls}

\end{document}